\documentclass[aps,prd,floats,preprint,tightenlines,nofootinbib]{revtex4}
\usepackage{rotating}
\usepackage{graphicx}
\begin{document}
\preprint{ \hbox{hep-ph/0412166} }
\vspace*{3cm}

\title{Effective Theory Analysis of Precision Electroweak Data}

\author{
        Zhenyu Han\footnote{email address:  {\tt zhenyu.han@yale.edu}}
        and  Witold Skiba\footnote{email address:  {\tt witold.skiba@yale.edu}}}
\affiliation{        \small \sl  Department of Physics, Yale University,
                New Haven, CT  06520\vspace{2.5cm} }

\begin{abstract}
We obtain the bounds on arbitrary linear combinations of operators of dimension 6
in the Standard Model. We consider a set of 21 flavor and CP conserving operators. Each of
our 21 operators  is tightly constrained by the standard set of electroweak
measurements. We perform a fit to all relevant precision electroweak data
and include neutrino scattering experiments, atomic parity violation, 
W mass, LEP1, SLD, and LEP2 data. Our results provide an efficient way of
obtaining bounds on weakly coupled extensions of the Standard Model. 
\end{abstract}

\maketitle

\newpage

 \section{Introduction}
Despite the enormous success of the Standard Model (SM), we are certain that the SM
is an effective theory with a cutoff  that is much smaller than the Planck scale.
 A lot of effort is being devoted to constructing and studying extensions
 of the SM that predict new particles having TeV scale masses.
 An integral part of this effort is constraining the parameters of new models
 using experimental data. In many cases, the constraints are obtained by
 directly computing the deviations from the SM that are induced by new particles
 in a specific model.

 An effective field theory approach is a two step process. First, one integrates
 out all new heavy states and obtains effective interactions involving only fields of the SM.
 These new effective higher-dimensional operators are then used to compute the deviations
 from the SM  and compare with the experimental data. Following this
 approach one needs to make contact with experimental data only
 once---by computing the effects of higher-dimensional operators on different experiments and
obtaining bounds on the coefficients of such operators using the data. Once this
step is completed, one can constrain any model just by calculating the coefficients
of new effective operators.

The effective theory approach is by no means new, and has been applied to the 
electroweak data many times. 
Lucid explanation on applying  effective Lagrangians to precision electroweak
measurements  can be found, for example, in  Refs.~\cite{Buchmuller,GW}. Perhaps the best
known example of the effective approach are the so-called oblique corrections~\cite{oblique,STU}. 
The oblique corrections only modify the SM gauge boson propagators. The formalism
of oblique corrections has been extended in Refs.~\cite{beyondSTU,LEP12}.
While in many models new physics contributions are limited to two point functions of gauge bosons, 
it is not always the case.

In this article we study bounds on the coefficients of effective operators that could be
relevant for the physics of electroweak symmetry breaking. We assume that just above the
electroweak scale the $SU(2)_L\times U(1)_Y$ gauge symmetry is linearly realized and
therefore the field content  involves a scalar electroweak doublet.
This determines the power counting and our  effective Lagrangian
 \begin{equation}
 \label{eq:Lagrangian}
   {\mathcal L}=  {\mathcal L}_{SM} + a_i \, O_i
 \end{equation}
contains operators of dimension 6 in addition to the SM Lagrangian. Therefore
each coefficient $a_i$ has the dimension of inverse mass squared and can be conveniently
represented as  $a_i=\frac{1}{\Lambda_i^2}$. In our analysis, 
we include 21 operators---two of which correspond to the $S$ and $T$ parameters~\cite{STU}.
Our choice of operators $O_i$ is explained in the next section.
In summary, we choose operators that conserve $U(3)^5$ flavor symmetry of the SM,
as well as conserve CP. We also restrict ourselves to operators that  are stringently constrained
by the usual set of electroweak precision measurements. There is a number of dimension
6 operators that are flavor and CP conserving but the available data is not accurate enough
to place stringent bounds on the corresponding $\Lambda_i$'s. The bounds on the scales
$\Lambda_i$ of the 21 individual operators in our basis  are all about 1 TeV or higher.
In contrast, the bounds on flavor conserving four-quark contact interactions or operators
involving quarks and gluons are usually much lower. While it is exciting that operators suppressed
by relatively low scales $\Lambda$ are allowed, such operators are not very useful for
constraining new models.

Because of the spectacular agreement between the SM and precision experiments
the allowed deviations from the SM are small. Therefore, when computing the effects of operators $O_i$
in Eq.~(\ref{eq:Lagrangian}) we can restrict ourselves to computing only the interference
terms between ${\mathcal L}_{SM} $ and the operators $O_i$. In other words, we work to the
linear order in the coefficients $a_i$. Expansion in powers of $a_i$ corresponds to expansion
in $\frac{E^2}{\Lambda_i^2}$ or $\frac{v^2}{\Lambda_i^2}$, where $E$ is the characteristic
energy scale of a given process and $v$ is the Higgs vacuum expectation value (vev).
Given that $E,v \ll \Lambda_i$, neglecting the terms quadratic
in $a_i$ is a good approximation. The SM predictions can be computed
to arbitrary accuracy since they do not depend on $a_i$. We then compare the results with
the experimental data and compute the $\chi^2$ distribution as a function of $a_i$.
Since our results are linear in the coefficients $a_i$, $\chi^2$ is quadratic in $a_i$
and can be written as
\begin{equation}
\label{eq:main}
  \chi^2(a_i)=\chi^2_{min} + (a_i - \hat{a}_i)  {\mathcal M}_{ij} (a_j-\hat{a}_j),
\end{equation}
 where $\hat{a}_i$ are the values of $a_i$ that minimize $\chi^2$.

The main result of our paper is the matrix ${\mathcal M}_{ij}$ and the vector $\hat{a}_i$
in Eq.~(\ref{eq:main}). We would like to stress an obvious point here---the coefficients
${\mathcal M}_{ij}$ and $\hat{a}_i$ are constants that we obtain by fitting to experimental data.
Given this, admittedly, large set of numbers one can compute
the bounds on any linear combination of operators $O_i$ in our basis.
Integrating out heavy particles from an  extension of the SM will, in general,
lead to a set of operators $O_i$ whose coefficients are correlated by the underlying theory.
Since we are working in the linear approximation in terms of the coefficients $a_i$, the effects
of correlated operators are also linear and the analysis of bounds on a new model
remains very simple. Readers interested in applying our work to their favorite model
can use  Eq.~(\ref{eq:main}) and skip all the details on how we obtain the $\chi^2$ distribution.

An effective Lagrangian analysis has been performed in many cases in the past.
For examples see Refs.~\cite{Buchmuller,GW,CGB,
Appelquist:1993ka,BarbieriStrumia,BGKLM,Arzt:1994gp}.
Ref.~\cite{BarbieriStrumia} is perhaps the closest to our 
approach, but it only includes the operators constrained by the Z-pole measurements
and considers bounds on individual operators.
We focus on a broader set of effective operators and
use the results from LEP1, SLD, LEP2, as well as lower energy experiments to bound
the coefficients of the effective operators. The most important set of precision
electroweak measurements come from LEP and SLD. Since these experiments
no longer collect data, the precision of the data relevant for our analysis is not going
to significantly improve in the near future.

In the next section we enumerate the operators in our basis. We further explain what motivates
our choice of the 21 operators.  In Sec.~\ref{sec:Experiments},  we list the experiments that
we use to obtain the bounds on the effective operators. We outline our calculations of the
interference terms between the SM and the additional operators in Sec.~\ref{sec:Calculations},
and describe the numerical analysis in Sec.~\ref{sec:chi2}.
We summarize the results in Sec.~\ref{sec:Summary}. 
Our numerical results are presented in Appendix~\ref{app:Matrix}. We show how to
use these results in Appendix~\ref{app:ST} by reproducing the bounds on the $S$ and
$T$ parameters. Another example is provided in Appendix~\ref{app:Zprime}, where
we illustrate our procedure on $Z'$ gauge bosons and compare our results
with Ref.~\cite{Zprime}. We performed the numerical analysis using Mathematica, and we make
the notebooks available online. A few comments about our code are contained 
in Appendix~\ref{app:program}.

 \section{Operators}
 \label{sec:Operators}
We assume that just above the electroweak symmetry breaking scale, the effective
theory is that of the SM with one Higgs doublet.
In our effective theory the $SU(2)_L \times U(1)_Y$ symmetry is linearly realized.
The assumption of one Higgs doublet is not at all relevant since
it is  only the vev of the Higgs that enters the analysis.
Since the electroweak symmetry is linearly realized the physical Higgs and the
electroweak breaking  vev are assigned mass dimension one.
The vev appearing in the Lagrangian is always raised to positive powers.
In case of nonlinearly realized electroweak symmetry, the expansion
parameter is the momentum divided by the vev times $4 \pi$.
A number of authors studied electroweak precision data in nonlinear
realizations of electroweak symmetry,
for examples see Refs.~\cite{Longhitano,Appelquist:1993ka,lectures}.

A complete set of dimension-6 operators consistent with
the $SU(3)\times  SU(2)_L \times U(1)_Y$, baryon and lepton number conservation
and the linearly realized electroweak symmetry has been presented in Ref.~\cite{Buchmuller}.
There are 80 operators in the basis of Ref.~\cite{Buchmuller}
after the leading order equations of motion are used
to obtain an independent set of operators~\cite{Politzer,Arzt:1993gz}.

We are interested in constraining models of new physics pertinent to the electroweak
symmetry breaking. Processes that contribute to flavor or CP violation have to  be
suppressed by  scales much higher than the electroweak scale. A typical suppression for
 four fermion operators that contribute to the $K-\bar{K}$ mass difference is about 
$10^3~{\rm TeV}$~\cite{Buchmuller}. The bounds are even more stringent in the lepton
sector: about $10^4~{\rm TeV}$ suppression is required for the electric dipole moment
of the electron~\cite{Regan:2002ta,Bernabeu:2004ww}; the limits on the $\mu\rightarrow e\gamma$ 
decay also imply $10^4~{\rm TeV}$ suppression for the contributing operators~\cite{Buchmuller}.
Thus it is natural to assume that the electroweak scale and the scales associated with flavor
and CP violation are well separated.  Processes involving the third generation are an
obvious exception. Experiments have limited statistics and new flavor-dependent physics
at TeV scale is possible. It is plausible that the third generation actively participates
in the electroweak symmetry breaking. This interesting case merits a separate study 
but is beyond the scope of this article. We therefore impose $U(3)^5$ flavor symmetry
on our operators. A different $U(3)$ acts on the  left-handed quarks and leptons as well as on
the right handed quarks and leptons. Consequently, our operators are
unchanged when written in terms of either mass or gauge eigenstates.

Among the 80 operators of dimension 6 listed in Ref.~\cite{Buchmuller}, there are 28 operators
that do not conserve CP or flavor $U(3)^5$, or violate both. Among the remaining 52 operators
there are 18 operators that only involve quark and/or gluon fields. The bounds on such operators
are poor since the precision of hadron experiments is not comparable to that of the $e^+ e^-$ machines.
Such operators are therefore not helpful in constraining models with new physics at the TeV scale.

Thus, there are 34 operators that conserve flavor and CP and contain either electroweak gauge
bosons or leptons and perhaps some quark fields. Six of these 34 operators are not observable
in current experiments: either they contribute to dimension 4 couplings in the SM
Lagrangian or they involve the Higgs doublet only.

We are finally left with 28 operators. Seven operators in this set are of the form
\begin{equation}
\label{eq:GammaoverMZ}
 O_{fF}= i (\bar{f} \gamma^\mu D^\nu f) \, F_{\mu \nu},
\end{equation}
where $f$ represents a fermion and $F_{\mu \nu}$ is the  field strength for
the hypercharge or weak gauge bosons. These operators can only contribute to the
Z-pole measurements. At other energies the interference term between the SM contribution and
the contribution of operators  $O_{fF}$ vanishes since one of the contributions is real
and the other is imaginary. For the same reason the interference term at the Z pole
is suppressed by the ratio of the Z width to the Z mass. 
We therefore neglect the operators $O_{fF}$ in our analysis.

We choose the following basis for the remaining 21 operators that are the focus of our work.
Our notation is standard: $q$ and $l$ represent the three families of the
left-handed quark and lepton fields, respectively. The right handed fields are labeled $u$, $d$, and $e$.
We omit the family index which is always summed over due to the flavor $U(3)^5$ symmetry.
We adopt the notation of Ref.~\cite{Buchmuller} with minor modifications
and  in a few cases our operators do differ from Ref.~\cite{Buchmuller} by a numerical factor.

The operators that contain only the gauge bosons and Higgs doublets are
\begin{equation}
\label{eq:owbh}
 O_{W\!B}=(h^\dagger \sigma^a h) W^a_{\mu \nu} B^{\mu \nu}, \  \  \   O_h = | h^\dagger D_\mu h|^2,
\end{equation}
where $W^a_{\mu \nu}$ is the $SU(2)$ field strength, $B_{\mu \nu}$ the hypercharge field strength,
and $h$ represents the Higgs doublet.
There are 11 four-fermion operators. These are
\begin{eqnarray} &&
  O_{ll}^s=\frac{1}{2} (\overline{l} \gamma^\mu l) (\overline{l} \gamma_\mu l), \ \ \
  O_{ll}^t=\frac{1}{2} (\overline{l} \gamma^\mu \sigma^a l) (\overline{l} \gamma_\mu \sigma^a l),
      \label{eq:oll} \\ &&
  O_{lq}^s= (\overline{l} \gamma^\mu l) (\overline{q} \gamma_\mu q), \ \ \
  O_{lq}^t= (\overline{l} \gamma^\mu \sigma^a l) (\overline{q} \gamma_\mu \sigma^a q),
     \label{eq:olq} \\ &&
  O_{le}= (\overline{l} \gamma^\mu l) (\overline{e} \gamma_\mu e),  \ \ \
  O_{qe}=(\overline{q} \gamma^\mu q) (\overline{e} \gamma_\mu e),
     \label{eq:olqe}  \\ &&
  O_{lu}= (\overline{l} \gamma^\mu l) (\overline{u} \gamma_\mu u),  \ \ \
  O_{ld}= (\overline{l} \gamma^\mu l) (\overline{d} \gamma_\mu d),
      \label{eq:olud} \\ &&
  O_{ee}=\frac{1}{2} (\overline{e} \gamma^\mu e) (\overline{e} \gamma_\mu e), \ \ \
  O_{eu}=(\overline{e} \gamma^\mu e) (\overline{u} \gamma_\mu u),  \ \ \
  O_{ed}=(\overline{e} \gamma^\mu e) (\overline{d} \gamma_\mu d).
      \label{eq:oeeud}
  \end{eqnarray}
  The operators in Eqs.~(\ref{eq:oll}) and (\ref{eq:olq}) involve only left-handed fields,
  in Eqs.~(\ref{eq:olqe}) and (\ref{eq:olud}) 2 left-handed and 2 right-handed,
  while  in Eq.~(\ref{eq:oeeud})  all right-handed fields. There are 7 operators containing
  2 fermions that alter the couplings of fermions to the gauge bosons
 \begin{eqnarray} &&
   O_{hl}^s = i (h^\dagger D^\mu h)(\overline{l} \gamma_\mu l) + {\rm h.c.}, \ \ \
   O_{hl}^t = i (h^\dagger \sigma^a D^\mu h)(\overline{l} \gamma_\mu \sigma^a l)+ {\rm h.c.},
     \label{eq:ohl} \\ &&
   O_{hq}^s = i (h^\dagger D^\mu h)(\overline{q} \gamma_\mu q)+ {\rm h.c.}, \ \ \
   O_{hq}^t = i (h^\dagger \sigma^a D^\mu h)(\overline{q} \gamma_\mu \sigma^a q)+ {\rm h.c.},
    \label{eq:ohq} \\ &&
      O_{hu} = i (h^\dagger D^\mu h)(\overline{u} \gamma_\mu u)+ {\rm h.c.}, \ \ \
   O_{hd} = i (h^\dagger D^\mu h)(\overline{d} \gamma_\mu d)+ {\rm h.c.},
           \label{eq:Ohud}  \\ &&
   O_{he} = i (h^\dagger D^\mu h)(\overline{e} \gamma_\mu e)+ {\rm h.c.}\,.
     \label{eq:ohe}
\end{eqnarray}
Finally, there is an operator that  modifies the triple gauge  boson interactions
 \begin{equation}
 \label{eq:ow}
  O_W = \epsilon^{abc} \, W^{a \nu}_{\mu} W^{b\lambda}_{\nu} W^{c \mu}_{\lambda}.
\end{equation}
Eqs.~(\ref{eq:owbh}) through (\ref{eq:ow}) define our basis of the 21 operators.

We denote the coefficients $a_i$ in the Lagrangian in Eq.~(\ref{eq:Lagrangian}) using the same
indices as the corresponding operators, so that the effective Lagrangian is
 \begin{equation}
 \label{eq:effLagrangian}
   {\mathcal L}=  {\mathcal L}_{SM} + a_{W\!B} \, O_{W\!B} + a_h \, O_h + \ldots + a_W\,   O_W.
 \end{equation}
 Note that the first two operators in our basis, $O_{W\!B} $ and $O_h$, are in a one-to-one
relation with the $S$ and $T$ parameters~\cite{STU}. The correspondence is
 \begin{equation}
 \label{eq:ST}
  a_{W\!B}= \frac{1}{4 s c} \frac{\alpha}{v^2} S, \\\ a_h = - 2 \frac{ \alpha}{v^2} T,
 \end{equation}
 where $\langle h \rangle =\left( \begin{array}{c}  0 \\ v/\sqrt{2} \end{array} \right)$,
 $\alpha$ is the fine-structure constant, $s$ and $c$ are the sine and cosine
 of the weak mixing angle, respectively.
 The $U$ parameter is related to a dimension-8 operator in our power counting scheme
 and therefore does not appear in our analysis.

 \section{Experiments}
  \label{sec:Experiments}

The three most precisely measured electroweak sector observables: $\alpha$, $G_F$,
and $M_Z$ are taken to be the input parameters, from which the SM gauge couplings
and the Higgs vev are inferred. Predictions for experiments are
computed in terms of the inputs and the coefficients of the new operators.
 The experimental quantities we use to constrain the coefficients of operators are listed in
Table~\ref{table:experiments}. Detailed descriptions and references for individual 
experiments can be found in many reviews, for example in Refs.~\cite{erler+langacker} 
and \cite{LEPII}.

\begin{table}[htb]
\begin{tabular}{|c|c|c|c|}
\hline
 & Standard Notation &  Measurement &Reference\\
 \hline
 Atomic parity &$Q_W(Cs)$&Weak charge in Cs&\cite{QWCs}\\
 violation& $Q_W(Tl)$& Weak charge in Tl&\cite{QWTl}\\
 \hline
   DIS&$g_L^2,g_R^2$&$\nu_\mu$-nucleon scattering from NuTeV&\cite{NuTeV}\\
      & $R^\nu$&$\nu_\mu$-nucleon scattering from CDHS and CHARM&\cite{CDHS,CHARM}\\
      &$\kappa$&$\nu_\mu$-nucleon scattering from CCFR&\cite{CCFR}\\
      &$g_V^{\nu e},g_A^{\nu e}$&$\nu$-$e$ scattering from CHARM II&\cite{CHARMII}\\

 \hline
   Z-pole&$\Gamma_Z$&Total $Z$ width&\cite{LEPII}\\
         &$\sigma_h^0$ &$e^+e^-$ hadronic cross section at $Z$ pole&\cite{LEPII}\\
         &$R_f^0(f=e,\mu,\tau,b,c)$& Ratios of decay rates &\cite{LEPII}\\
         &$A_{FB}^{0,f}(f=e,\mu,\tau,b,c)$ &Forward-backward asymmetries&\cite{LEPII}\\
         &$\sin^2\theta_{eff}^{lept}(Q_{FB})$&Hadronic charge asymmetry&\cite{LEPII}\\
         &$A_f(f=e,\mu,\tau,b,c)$&Polarized asymmetries&\cite{LEPII}\\
 \hline
 Fermion pair    &$\sigma_f(f=q,\mu,\tau)$& Total cross sections for 
$e^+e^-\rightarrow f\overline f$&\cite{LEPII}\\
 production at   &$A_{FB}^f(f=\mu,\tau)$& Forward-backward asymmetries for 
$e^+e^-\rightarrow f\overline f$&\cite{LEPII}\\
 LEP2            &$d\sigma_e/d\cos\theta$&Differential cross section for 
$e^+e^-\rightarrow e^+e^-$&\cite{OPALfpair}\\
 \hline
 $W$ pair&$d\sigma_W/d\cos\theta$& Differential cross section for $e^+e^-\rightarrow W^+W^-$&\cite{L3Wpair}\\
 \hline
 &$M_W$ & W mass &\cite{LEPII,WmassTevatron}\\

 \hline
\end{tabular}
\caption{\label{table:experiments} Relevant measurements}
\end{table}

The list of experiments in Table~\ref{table:experiments} does not include
the anomalous magnetic moment of the muon~\cite{gminus2}, one of the 
most precisely measured electroweak quantities. The operators that contribute
directly to $(g-2)$ involve left and right-handed fields and are not $U(3)^5$ invariant. 
There are also loop contributions from operators like $O_{W\!B}$, $O_W$,
and many four-fermion operators. Such loop contributions are divergent and
require introducing  counterterms in the form of  operators excluded from our analysis
due to their lack of $U(3)^5$ invariance. An operator analysis of contributions to the
muon $(g-2)$ can be found in Ref.~\cite{AEW}.

For a given observable $X$, our prediction can be written as:
\begin{equation}
\label{prediction} 
   X_{th}=X_{SM}+\sum_i a_i X_i,
\end{equation}
where $X_{th}$ is the prediction in the presence of additional operators, 
$X_{SM}$ is the standard model prediction and $\sum_ia_iX_i$
are corrections from our new operators. In practice, the SM predictions
are computed to the required accuracy in perturbation theory
and are well known for all the measurements we use.
Note that the corrections $X_i$  arise in two different ways. 
First, an operator can generate a new Feynman diagram contributing 
to a given physical process.  For example, a four-fermion operator
$O_{le}$ enters the $e^+e^-\rightarrow \mu^+\mu^-$ process as a new diagram, 
in addition to the $Z$ and $\gamma$ exchange
diagrams. We call this ``direct'' correction. Second, some operators
can shift the input parameters, because they add new
diagrams to the physical processes based on which $\alpha$, 
$G_F$, and $M_Z$ are measured. Thus, the input parameters 
determined from these observables are different from their SM values.
Since all the other observables are calculated from these
input parameters, they will inevitably receive indirect corrections
from the shifts. We summarize the direct and indirect effects of our operators 
in Table~\ref{table:operators}.

Because of their high statistics, the Z-pole data and several best
measured low-energy observables dominate the bounds on the coefficients $a_i$
whenever such measurements constrain an operator. This is the case for the
operators that shift the input parameters and the operators of the
form $O_{hf}$, which change the couplings between the $Z$ boson and
the fermions. The four-fermion operators do not contribute to
the $Z$-pole measurements at the linear order, 
that is the interference term between the SM contribution
and  the four-fermion operators vanishes at the Z pole.
Therefore, to constrain four-fermion operators we have to include the 
cross sections for fermion-pair production at LEP2. We also include the differential
cross sections for the $W$ pair production to constrain the operator
$O_W$. There are several operators, in addition to $O_W$, that alter the cross section
for $W$ pair production. However, these operators are well bounded by other
measurements and the $W$ pair production does not contribute significantly to
the bounds on their coefficients.

\begin{table}[hbt]
\begin{tabular}{|c|c|c|c|c|c|c|c|}
\hline
 Operator(s)&shift&$M_W$&Z-pole&DIS&$ Q_W$&$e^+e^-\rightarrow f\overline f$
 (LEP2)&$e^+e^-\rightarrow W^+W^-$\\
 \hline
 $O_{W\!B}$&$\alpha, \, M_Z$& &$\surd$&$\surd$&$\surd$&$\surd$&$\surd$\\
 \hline
 $O_h$&$M_Z$&&&&&&\\
 \hline
 $O_{ll}^t$&$G_F$&&&$\surd$&&$\surd$&\\
 \hline
 $O_{ll}^s$, $O_{le}$ &&&&$\surd$&&$\surd$&\\ \hline
 $O_{ee}$ & & & & & & $\surd$ & \\ \hline
 $O_{lq}^s,O_{lq}^t,O_{lu},O_{ld}$&&&&$\surd$&$\surd$&$\surd$&\\ \hline
 $O_{eq},O_{eu},O_{ed}$&&&&&$\surd$&$\surd$&\\
 \hline
 $O_{hl}^t$&$G_F$&&$\surd$&$\surd$&$\surd$&$\surd$&$\surd$\\
 \hline
 $O_{hl}^s,O_{he}$&&&$\surd$&$\surd$&$\surd$&$\surd$&$\surd$\\
 \hline
 $O_{hu},O_{hd},O_{hq}^s,O_{hq}^t$&&&$\surd$&$\surd$&$\surd$&$\surd$&\\
 \hline
 $O_W$&&&&&&&$\surd$\\
\hline
\end{tabular}
\caption{ \label{table:operators}
Measurements influenced by different operators.
The check marks, $\surd$,  indicate  ``direct" corrections only. When an operator
contributes to one of the input parameters, the corresponding shift of the input parameter 
does affect all measurements. }
\end{table}

\section{Calculations}
\label{sec:Calculations}
In this section we describe the computation of the effects of 
dimension-6 operators. Since a lot of work on this topic is already
available in the literature, we quote the results whenever available.
We have independently verified all the quoted results.

We work in the linear approximation
in terms of the coefficients $a_i$. As we indicated in the previous
section, there are two ways that terms linear in $a_i$ arise.
First, as a result of additional Feynman diagrams due to the
dimension-6 operators. In this case we simply compute the
interference terms between the new operators and the tree-level
contribution in the SM. Second, a few of our operators 
redefine the input parameters inferred from the measurements
of $\alpha$, $M_Z$, and $G_F$. We use the tree-level SM results
and expand to the linear order in the deviations induced by the
coefficients $a_i$. One of the most transparent ways of dealing with
the shifts of the input parameters is described in Ref.~\cite{BGKLM}.

To track down all the shifts, we
find it convenient to use the following parameters in the SM
Lagrangian: $e$, $s$, and $M_Z$; or equivalently $g$, $g'$, and $M_Z$. 
They are related to the input
parameters and the new operators as
\begin{eqnarray}
e^2&=&e^2_0 (1-2v^2s c \, a_{W\!B}),\label{Sshift}\\
M_Z^2&=&M_{Z0}^2(1+2v^2sc\, a_{W\!B}+ \frac{v^2 }{2 }a_h),\label{STshift}\\
\frac{1}{v^2}&=&\frac{1}{v_0^2}+ 2 a_{hl}^t -a_{ll}^t,
\end{eqnarray}
where
\begin{equation}
\alpha=\frac{e_0^2}{4\pi},\quad
G_{F}=\frac{1}{\sqrt{2}v_0^2}=\frac{e^2_0}{4 \sqrt{2} s_0^2 c_0^2 M_{Z0}^2}.
\end{equation}
Parameters with subscripts $0$ are the values derived in the absence
of any additional operators.
Eqs.~(\ref{Sshift}) and (\ref{STshift}) are the shifts
due to the $S$ and $T$ parameters. 

The corrections induced by $S$ and $T$ are given in Ref.~\cite{STU}, 
and  can be easily translated to our notation using Eq.~(\ref{eq:ST}). 
Ref.~\cite{STU} does not provide formulas for LEP2, but the extension to 
LEP2 is simple. The operator $O_{W\!B}$ also contains a triple gauge boson coupling, 
which contributes to the $e^+ e^- \rightarrow W^+ W^-$ process. We discuss this
at the end of this section.

 The operators with 2 Higgs doublets and 2 fermions, $O_{hf}$, alter
the couplings between gauge bosons and fermions. The changes to the
$Z$-fermion couplings,  $g_V^f$ and $g_A^f$, are given in
Ref.~\cite{BarbieriStrumia}. (The $v^2$ in Ref.~\cite{BarbieriStrumia}
is defined as one half of our value.) These changes affect all
measurements involving a $Z$-exchange diagram. We have
calculated the SM tree-level predictions in terms of arbitrary
$g_V^f$ and $g_A^f$, so it is easy to obtain the corrections induced by $O_{hf}$. 
In addition, the operators $O_{hl}^t$ and $O_{hq}^t$ change the couplings of
the $W$ boson to the leptons and quarks:
\begin{eqnarray}
 g&\rightarrow&g(1+v^2
       a_{hl}^t)\qquad(W-\mbox{leptons coupling}),\\
 g&\rightarrow&g(1+v^2
 a_{hq}^t)\qquad(W-\mbox{quarks coupling}).
\end{eqnarray}
These changes enter, besides the shift to $G_F$, the calculations of
the $e^+e^-\rightarrow W^+W^-$ cross section in the $\nu$ exchange
channel, and the $\nu$-nucleon scattering charged current
cross sections.

The operator $O_{ll}^t$ shifts the value of the input parameter inferred from 
$G_F$. All other four-fermion operators do not contribute to the $Z$-pole measurements.
However, they contribute to the low-energy measurements and LEP2 measurements.
 We now enumerate the effects of four-fermion operators on the relevant  observables:
\begin{enumerate}
 \item The weak charges measured in atomic parity violation
 experiments
 \begin{equation}
 Q_W(Z,N)=-2[(2Z+N)C_{1u}+(Z+2N)C_{1d}].
 \end{equation}
 $C_{1u}$ and $C_{1d}$ receive corrections
 \begin{eqnarray}
 \Delta 
C_{1u}&=&\frac{\sqrt{2}}{4G_F}(-a_{lq}^s+a_{lq}^t+a_{eu}+a_{qe}-a_{lu}),\\
 \Delta 
C_{1d}&=&\frac{\sqrt{2}}{4G_F}(-a_{lq}^s-a_{lq}^t+a_{ed}+a_{qe}-a_{ld}).
 \end{eqnarray}
\item $\nu$-nucleon scattering. The 4-fermion operators affect both the 
neutral
 current and the charged current cross sections. The corrections to the
couplings $g_L^u, g_L^d,g_R^u,g_R^d$ are
 \begin{eqnarray}
 \Delta g_{L,eff}^u&=&\frac{1}{2\sqrt{2}G_F} [- a_{lq}^s + (2 g_L^u-1) a_{lq}^t],\\
 \Delta g_{L,eff}^d&=&\frac{1}{2\sqrt{2}G_F} [- a_{lq}^s+ (2 g_L^d+1) a_{lq}^t],\\
 \Delta g_{R,eff}^u&=&\frac{1}{2\sqrt{2}G_F} (- a_{lu}^s+2 g_R^u a_{lq}^t),\\
 \Delta g_{R,eff}^d&=&\frac{1}{2\sqrt{2}G_F} (- a_{ld}^s+2 g_R^d a_{lq}^t).
\end{eqnarray}
 These corrections are``effective" in the sense that they are not
corrections to the $Z$-fermion couplings and thus the formulas above only
apply to the $\nu$-nucleon scattering. The corrections to the measured 
quantities  can be easily calculated from the above equations, for example,
 $g_L^2=(g_L^u)^2+(g_L^d)^2$ measured at NuTeV receive the correction
 \begin{equation}
 \Delta(g_L^2)=2g_L^u \Delta g_{L,eff}^u+2g_L^d\Delta g_{L,eff}^d.
 \end{equation}
\item $\nu$-$e$ scattering. The corrections to the coupling
 $g_V^{\nu e}$ and $g_A^{\nu e}$ are
 \begin{eqnarray}
 \Delta g_{V,eff}^{\nu 
e}&=&\frac{1}{2\sqrt{2}G_F} (-a_{ll}^s+a_{ll}^t -a_{le}),\\
 \Delta g_{A^eff}^{\nu e}&=&\frac{1}{2\sqrt{2}G_F} (-a_{ll}^s+a_{ll}^t+a_{le}).
 \end{eqnarray}
 Again, these corrections are ``effective'' that is only apply to the
 $\nu$-$e$ scattering process.
 \item Fermion pair production at LEP2 energies. The differential
 cross sections in the presence of the contact operators are given,
 for example, in Ref.~\cite{Kroha:1991mn}.
\end{enumerate}

 Finally, the operators $O_W$ and $O_{W\!B}$ alter the
 triple gauge boson couplings.  After substituting the vev for the Higgs doublet,
 the two operators yield the following couplings 
 \begin{equation}
 \label{eq:3gauge}
\Delta \mathcal{L}= ia_{W\!B} v^2 gW^+_\mu W^-_\nu(c A^{\mu\nu}-s
  Z^{\mu\nu})+6ia_WW^{-\mu}_\nu
  W^{+\lambda}_\mu(sA^\nu_\lambda+cZ_\lambda^\nu).
 \end{equation}
The tree-level $e^+e^-\rightarrow W^+W^-$
differential cross section is calculated in
Ref.~\cite{triplegauge} for arbitrary triple gauge boson couplings. 
Our effective couplings, Eq.~(\ref{eq:3gauge}), correspond to the terms multiplying
 $\kappa_V$ and $\lambda_V$ in Eq.~(2.1) in Ref.~\cite{triplegauge}.
 To obtain the  cross section we substitute
\begin{eqnarray}
 \Delta\kappa_\gamma&=&\frac{ v^2 c}{s} \,a_{W\!B} ,\\
 \Delta\kappa_Z&=&-\frac{v^2s}{c} \, a_{W\!B},\\
 \Delta\lambda_\gamma&=&\Delta\lambda_Z=\frac{3v^2 g}{2} \,a_W .
\end{eqnarray}
where $\Delta$'s denote the deviations from the SM values.

\section{Total $\chi^2$ distribution}
\label{sec:chi2}
In the previous section, we described how to compute
the changes of observable quantities induced by the operators.
Given theses results we calculate the total $\chi^2$ distribution. For non-correlated measurements,
\begin{equation}
\label{eq:chi2uncorr} 
  \chi^2(a_i)=\sum_X\frac{(X_{th}(a_i)-X_{exp})^2}{\sigma_X^2},
\end{equation}
where $X_{exp}$ is the experimental value for observable $X$ and
$\sigma_X$ is the total error both experimental and theoretical. The
experimental values for the observables are obtained from the
references cited in Table~\ref{table:experiments}.
The input parameters and the SM
 predictions, except those for LEP2, are obtained from
Ref.~\cite{erler+langacker}, where the following input parameters
are used:
\begin{equation}
\label{eq:inputs}  m_{Higgs}=113~\mbox{GeV},\ 
m_{top}=176.9~\mbox{GeV},\ \alpha_s(M_Z)=0.1213
\end{equation}
in addition to $M_Z$, $\alpha$, and $G_F$. The uncertainty in the values of the input parameters
are incorporated as theory errors on the SM predictions and combined later
with experimental errors.
 For the DIS measurements of CDHS, CHARM and CCFR, we use the SM
predictions in Ref.~\cite{Erler+Langacker2002}, but have corrected
them for the small differences of input parameters~\cite{Langacker:1991zr}.
The sensitivities of the SM predictions for the $e^+e^-\rightarrow f^+f^- (f\neq e)$
cross sections at LEP2 and the $e^+e^-\rightarrow W^+W^-$ cross section
to small changes of the input parameters are negligible compared to experimental 
errors, as we have verified using ZFITTER~\cite{zfitter} and RacoonWW~\cite{RacoonWW}. 
Therefore, we use the SM predictions provided in the corresponding experimental references.
The SM prediction for the $e^+e^-\rightarrow e^+e^-$ (LEP2) differential
cross section is calculated using the program BHWIDE~\cite{BHWIDE},
assuming the same input parameters as in Eq.~(\ref{eq:inputs}).

The SM predictions agree with the experimental values
well, except for a  significant discrepancy for
$g_L^2$ obtained by the NuTeV collaboration~\cite{NuTeV}.
We include the NuTeV result in our calculation, but
one could easily omit this result from the  $\chi^2$ calculation.
We also note that the LEP2 results for the
$e^+e^-\rightarrow q\overline q$ total cross sections are
larger than the SM predictions fairly consistently across different
energies probed by LEP2. If we tried to constrain the coefficients
of four-fermion operators with two leptons and two quarks using
only the  $e^+e^-\rightarrow q\overline q$ total cross sections, 
we would get relatively weak bounds on the
coefficients $a_i$ of such operators. Weak bounds mean
that the the contributions quadratic in the corrections in
$a_i$'s should not be neglected, contrary to what we do.
However, this discrepancy is not supported by other data that also constrains the 
same operators. A combined fit to all observables does not yield any coefficients
$a_i$ large enough to invalidate the linear approximation.
Therefore we suspect this pattern is caused by a  systematic error.

Higher order terms in perturbation theory alter our tree-level
calculation of the interference terms between the SM and the
additional operators. The most important effect in electron-positron
scattering is the initial state QED radiation. In order to assess
the effect of the radiative corrections, we compare the bounds on
the 4-fermion contact operators  for the $e^+e^-\rightarrow
l^+l^-(l\neq e)$ and $e^+e^-\rightarrow q\overline q$ channels with
the results given in Ref.~\cite {LEPII}, Table 8.13. In Ref.~\cite{LEPII}, the $e^+e^-
\rightarrow q\overline q$ channel is constrained using the inclusive
hadronic total cross sections. The $e^+e^- \rightarrow
l^+l^-(l\neq e)$ channel is constrained using the total cross
sections and asymmetries for $l=\mu,\tau$ and assuming equal
coefficients for the contact terms with $\mu$ and $\tau$. For the
purpose of comparison we use the same data sets and carry out the
same fits. In our final results,
 of course, all the data is taken into account to obtain
the bounds on contact operators.

The comparison was carried out for different contact interactions.
(The coefficients of 4-fermion operators in Ref.~\cite {LEPII}
differ from our coefficients $a_i$ by a factor
of $4 \pi$.) Several radiative corrections and second order terms in
$a_i$ are considered when obtaining bounds on these
coefficients\footnote{The radiative corrections are not mentioned in
Ref.~\cite{LEPII}, but are described in results of individual
experiments at LEP2. See, for example,
Refs.~\cite{OPALfpair}~\cite{DELPHIfpair}.}. We have obtained bounds
for the coefficients of operators neglecting radiative corrections
but including second order terms in $a_i$. Except for one case, the
differences between our bounds and the bounds in Ref.~\cite{LEPII}
for errors on $a_i$ are less than 20\%, and for centra values
of $a_i$ are less than $0.2\sigma$. When translated to the
scale $\Lambda$, the difference is less than 10\%, which is
satisfactory for our purposes. The exception is the bound for the
``LR" interaction for $e^+e^- \rightarrow q\overline q$ channel,
which corresponds to $a_{lu}$ and $a_{ld}$ in our notation, in which
case the errors quoted in Ref.~\cite{LEPII} are much more asymmetric
than our estimate of errors. However, the central value of the
coefficients differs less than $0.2\sigma$. Since the contact
terms with two leptons and two quarks can be constrained much more
stringently by the low energy measurements, this discrepancy 
cannot significantly affect our global fit.

For $e^+e^-\rightarrow f^+f^- \quad(f\neq e)$ channels, we have also
implemented the initial-state photonic correction to the order
$O(\alpha)$, which includes initial state soft photon exponentiation
and hard photon emission~\cite{zfitter}. This correction is the
largest radiative correction and we have obtained better agreement
with Ref.~\cite{LEPII} by including it. Except for the ``LR"
interaction in the  $e^+e^-\rightarrow q\overline q$ channel
mentioned previously, the discrepancies between our errors and
central values of $a_i$ and the results of \cite{LEPII} are within
10\% and $0.1\sigma$ respectively. The remaining discrepancy likely
arises from the final state radiative correction, pair production
correction and higher order corrections, that we have not
implemented.

We can safely neglect effects that contribute at a 10\% level
to our estimates of the coefficients $a_i$, we have nevertheless 
included the first order initial state QED corrections to
$e^+e^-\rightarrow f^+f^- \quad(f\neq e)$ channels in our
calculation of $\chi^2$. A factor of $(1+\alpha_s/\pi)$ is
also used to account for the QCD corrections for the hadronic final
states~\cite{OPALfpair130}.

Eq.~(\ref{eq:chi2uncorr}) must be modified to account for
the correlations between different measurements. There are three
categories of data for which the correlations between measurements cannot be
neglected. These are the correlations between $Z$-pole
observables~\cite{LEPII}, the experimental error correlations for the hadronic
total cross sections at LEP2 energies~\cite{LEPII}; and the theoretical
and experimental error correlations for $e^+e^-\rightarrow e^+e^-$
differential cross sections~\cite{Ward}. Including correlations,
\begin{equation}
 \label{eq:chi2corr}
  \chi^2(a_i)=\sum_{p,q}(X_{th}^p(a_i)-X_{exp}^p)(\sigma^2)^{-1}_{pq}(X_{th}^q(a_i)-X_{exp}^q),
\end{equation}
where the error matrix $\sigma^2$ is related to the standard deviations
$\sigma_p$ and the correlation matrix $\rho_{pq}$ as folows
\begin{equation}
 \sigma^2_{pq}=\sigma_p\rho_{pq}\sigma_q.
\end{equation}
Note that often the correlations for the theoretical, statistical and
systematic errors are different, and one should take this into account
when computing the final error matrix.

Numerical results for the $\chi^2$ distribution are presented in Appendix~\ref{app:Matrix}.

 \section{Summary}
  \label{sec:Summary}
We have obtained bounds on the coefficients of 21 dimension 6 operators in the SM\@.
Our analysis is linear in the coefficients of these operators. Therefore, the deviations from
the SM predictions arise as interference terms between the SM and the dimension 6 operators.
As is often the case, integrating out heavy particles leads not to just one but to several operators
whose coefficients are related in terms of the masses and coupling constants of the heavy states.
A global analysis of precision electroweak measurements must take into account all new operators
induced by integrating out heavy states and account for relations between the coefficients
of such operators. Our analysis allows obtaining bounds not just on each individual operator, but
on their linear combinations as well. Doing so, in the linear approximation, does not require
complicated numerical analysis, and can be done efficiently using our results. 
Of course, if the new physics contributions are ``oblique" or ``universal" only, \cite{STU,LEP12}, one
does not need the whole set of 21 operators. A subset of our operators that only modify SM
gauge boson propagators has already been considered in Refs.~\cite{STU,LEP12}

Our analysis is accomplished by computing the $\chi^2$ distribution as a function of the coefficients
$a_i$. In the linear approximation, $\chi^2$ takes the form shown in Eqs.~(\ref{eq:main})
and (\ref{eq:main2}). We concentrated on flavor and CP conserving operators.
Such operators are allowed in the SM when suppressed by scales of the order of a few TeV\@. 
Generic flavor and CP violating operators must be suppressed by much higher scales. This wide
separation of characteristic scales suggests that the electroweak symmetry breaking 
and the flavor and CP violating sectors can be analyzed independently of one another.
We excluded from our analysis operators that are not tightly constrained by the data,
for example operators involving only quarks and gluons. Such operators are not helpful
in constraining extensions of the SM.

The bounds on the coefficients of individual operators, by which we mean that the 
SM Lagrangian is amended by only one operator at a time, can be easily obtained
from Eq.~(\ref{eq:main}). The $1\, \sigma$ bound on a coefficient $a_k$ is $\hat{a}_k \pm 
\sqrt{{\mathcal M}_{kk}^{-1}}$, 
where ${\mathcal M}_{kk}$ indicates a diagonal element of ${\mathcal M}$
and is not summed over $k$. The fourth roots of the diagonal elements, $\sqrt[4]{{\mathcal M}_{kk}}$,
vary from $1.3~{\rm TeV}$ to  $17~{\rm TeV}$, which is a measure of how rapidly $\chi^2$ changes
as a function of $a_k=\frac{1}{\Lambda_k^2}$.

What is interesting is that the eigenvalues of ${\mathcal M}$ vary over a much
wider range, their fourth roots span from $180~{\rm GeV}$ to $21~{\rm TeV}$. 
In particular, the fourth roots of the four smallest eigenvalues are $180$, $250$, $390$
and $420~{\rm GeV}$. This means that four linear
combinations of the operators are much more weakly constrained than the individual
operators. The emergence of these ``weakly bounded directions" in the operator  space is an
interesting byproduct of our analysis. Of course, one can not trust the exact bounds on the 
``weakly bounded" operators. The linear analysis is not applicable when the suppression
scales are so low. One needs to work to the quadratic order in the coefficients to reliably
determine the bounds. However, it is clear that the bounds on such linear combinations
of operators are quite weak and below 1~TeV\@.   
It is interesting to find out if there are heavy particles that yield one of
the weakly constrained combination of operators when integrated out. This
possibility is currently being investigated.

\section*{Acknowledgments}
We thank Jens Erler, Alex Pomarol, and Pat Ward for helpful correspondences. We are
grateful to Tom Appelquist and Martin Schmaltz for discussions and comments.
This research was supported in part by the US Department of Energy
under grant  DE-FG02-92ER-40704. WS is also supported in part by the DOE OJI program.

\appendix
 \section{The Matrix}
 \label{app:Matrix}
Our main result can be presented in two alternative ways
\begin{equation}
\label{eq:main2}
  \chi^2=\chi^2_{min}+(a_i-\hat a_i){\mathcal M}_{ij} (a_j-\hat a_j)= \chi^2_{SM} + a_i \hat{v_i} +
          a_i {\mathcal M}_{ij} a_j.      
\end{equation}
In the equation above,
$\chi^2_{min}$ is the minimum of  $\chi^2$ in the presence of dimension 6 operators,
and $\chi^2_{SM}$ is the value of $\chi^2$ when all coefficients $a_i$ are zero.
The matrix ${\mathcal M}$ is symmetric and positive definite. The first equation 
makes apparent the values of coefficients $a_i$ that minimize $\chi^2$, which we call $\hat{a}_i$.
 The second equation is more convenient to use if only a few coefficients $a_i$ are not equal to zero.
\begin{table}[bht]
\begin{tabular}{c|ccccccccccc}
$a_i$ & $a_{W\!B}$ & $a_{h}$ & $a_{ll}^s$ & 
$a_{ll}^t$ & $a_{lq}^s$ & $a_{lq}^t$ & $a_{le}$ & 
$a_{qe}$ & $a_{lu}$ & $a_{ld}$ & $a_{ee}$ \\ \hline 
  $\hat{a}_i$ & $4.1\, 10^2$ & $-9.3\, 10^2$ & $-5.0$ & $-5.8$ & $-60.$ & $-6.9$ & 
$-0.3$ & $-23.$ & $-4.1\, 10^2$ & $-7.8\, 10^2$ & $7.5$ \\ 
$\hat{v}_i$ & $1.5\, 10^2$ & $-23.$ & $49.$ & $76.$ & $-1.1\, 10^2$ & $-2.4\, 10^2$ & 
$29.$ & $1.4\, 10^2$ & $-36.$ & $-68.$ & $44.$ \\ 
\hline \hline
 $a_i$ &$a_{eu}$ & $a_{ed}$ & $a_{hl}^s$ & $a_{hl}^t$ & 
$a_{hq}^s$ & $a_{hq}^t$ & $a_{hu}$ & $a_{hd}$ & 
$a_{he}$ & $a_{W}$  & \\ \hline
$\hat{a}_i$  &$-5.9\, 10^2$ & $-6.4\, 10^2$ & $2.3\, 10^2$ & $19.$ & $-77.$ & $14.$ & 
$-3.0\, 10^2$ & $98.$ & $4.6\, 10^2$ & $-4.4\, 10^2$ \\  
$\hat{v}_i$ & $1.0 \, 10^2$ & $15.$ & $-6.4\, 10^2$ & $-88.$ & $1.0\, 10^2$ & $1.7\, 10^2$ & 
$71.$ & $63.$ & $1.8\, 10^2$ & $1.0$ 
\end{tabular}
\caption{\label{table:av}
Coefficients $\hat{a}_i$ and $\hat{v}_i$ described in Eq.~(\ref{eq:main2}). To obtain values
of $\hat{a}_i$ one needs to multiply the numbers in the table times $10^{-8}~({\rm GeV})^{-2}$ and
to obtain $\hat{v}_i$ multiply times $10^{6}~({\rm GeV})^{2}$.}.
\end{table}

The two equivalent sets of coefficients $\hat{a}_i$ and $\hat{v_i}$ are presented in Table~\ref{table:av}.
The elements of matrix ${\mathcal M}$ are listed in Table~\ref{table:M}. The dimensions of these elements
are easy to read off Eq.~(\ref{eq:main2}) since $\chi^2$ is dimensionless and $a_i$'s have
dimension inverse mass squared.

\addtolength{\tabcolsep}{-0.5pt}
\begin{sidewaystable}
{\footnotesize
\begin{tabular}{c|ccccccccccccccccccccc}
$a_{W\!B}$ & $9.1e4$ &   &   & 
  &   &   &   & 
  &   &   &   & 
  &   &   &   & 
  &   &   &   & 
  &   \\[1pt] 
$a_{h}$ & $2.4e4$ & $7.9e3$ &   & 
  &   &   &   & 
  &   &   &   & 
  &   &   &   & 
  &   &   &   & 
  &   \\[1pt] 
$a_{ll}^s$ & $-78.$ & $-51.$ & $5.8e2$ & 
  &   &   &   & 
  &   &   &   & 
  &   &   &   & 
  &   &   &   & 
  &   \\[1pt] 
$a_{ll}^t$ & $-3.9e4$ & $-1.2e4$ & $6.7e2$ & 
$2.2e4$ &   &   &   & 
  &   &   &   & 
  &   &   &   & 
  &   &   &   & 
  &   \\[1pt] 
$a_{lq}^s$ & $-1.4e3$ & $-1.6e2$ & $0.$ & 
$1.5e2$ & $2.7e3$ &   &   & 
  &   &   &   & 
  &   &   &   & 
  &   &   &   & 
  &   \\[1pt] 
$a_{lq}^t$ & $-5.5e2$ & $-1.4e2$ & $0.$ & 
$5.9e2$ & $4.6e2$ & $2.9e3$ &   & 
  &   &   &   & 
  &   &   &   & 
  &   &   &   & 
  &   \\[1pt] 
$a_{le}$ & $-56.$ & $-9.7$ & $2.8e2$ & 
$3.0e2$ & $0.$ & $0.$ & $1.3e3$ & 
  &   &   &   & 
  &   &   &   & 
  &   &   &   & 
  &   \\[1pt] 
$a_{qe}$ & $1.3e3$ & $72.$ & $0.$ & 
$-1.4e2$ & $-2.7e3$ & $-7.4e2$ & $0.$ & 
$2.8e3$ &   &   &   & 
  &   &   &   & 
  &   &   &   & 
  &   \\[1pt] 
$a_{lu}$ & $-4.0e2$ & $3.8$ & $0.$ & 
$-1.1e2$ & $1.2e3$ & $-2.5e2$ & $0.$ & 
$-1.2e3$ & $7.1e2$ &   &   & 
  &   &   &   & 
  &   &   &   & 
  &   \\[1pt] 
$a_{ld}$ & $-6.9e2$ & $-6.9$ & $0.$ & 
$66.$ & $1.4e3$ & $3.3e2$ & $0.$ & 
$-1.4e3$ & $5.8e2$ & $7.8e2$ &   & 
  &   &   &   & 
  &   &   &   & 
  &   \\[1pt] 
$a_{ee}$ & $-59.$ & $-42.$ & $5.3e2$ & 
$6.1e2$ & $0.$ & $0.$ & $2.6e2$ & 
$0.$ & $0.$ & $0.$ & $4.8e2$ & 
  &   &   &   & 
  &   &   &   & 
  &   \\[1pt] 
$a_{eu}$ & $7.8e2$ & $1.1e2$ & $0.$ & 
$-2.1e2$ & $-1.3e3$ & $-9.1e2$ & $0.$ & 
$1.4e3$ & $-4.8e2$ & $-7.3e2$ & $0.$ & 
$8.4e2$ &   &   &   & 
  &   &   &   & 
  &   \\[1pt] 
$a_{ed}$ & $4.2e2$ & $-83.$ & $0.$ & 
$1.7e2$ & $-1.3e3$ & $5.5e2$ & $0.$ & 
$1.3e3$ & $-7.3e2$ & $-6.8e2$ & $0.$ & 
$4.7e2$ & $8.8e2$ &   &   & 
  &   &   &   & 
  &   \\[1pt] 
$a_{hl}^s$ & $-1.7e4$ & $-4.1e3$ & $1.5e2$ & 
$9.7e3$ & $-5.9e2$ & $8.3e2$ & $17.$ & 
$3.7e2$ & $-3.9e2$ & $-1.6e2$ & $1.3e2$ & 
$66.$ & $3.8e2$ & $5.5e4$ &   & 
  &   &   &   & 
  &   \\[1pt] 
$a_{hl}^t$ & $5.9e4$ & $1.7e4$ & $-43.$ & 
$-3.0e4$ & $-7.1e2$ & $-6.6e2$ & $-31.$ & 
$6.6e2$ & $-82.$ & $-3.4e2$ & $-32.$ & 
$4.9e2$ & $47.$ & $1.5e4$ & $6.3e4$ & 
  &   &   &   & 
  &   \\[1pt] 
$a_{hq}^s$ & $-1.9e3$ & $-1.4e3$ & $0.$ & 
$2.7e3$ & $-2.6e3$ & $-72.$ & $0.$ & 
$2.6e3$ & $-1.2e3$ & $-1.4e3$ & $0.$ & 
$1.2e3$ & $1.4e3$ & $-6.6e3$ & $-8.7e3$ & 
$6.0e3$ &   &   &   & 
  &   \\[1pt] 
$a_{hq}^t$ & $-9.3e3$ & $-4.5e3$ & $0.$ & 
$8.7e3$ & $-49.$ & $3.5e2$ & $0.$ & 
$56.$ & $-1.4e2$ & $-36.$ & $0.$ & 
$-64.$ & $1.8e2$ & $-2.4e4$ & $-3.1e4$ & 
$7.7e3$ & $2.6e4$ &   &   & 
  &   \\[1pt] 
$a_{hu}$ & $-6.1e2$ & $-6.6e2$ & $0.$ & 
$1.2e3$ & $-1.2e3$ & $-4.$ & $0.$ & 
$1.2e3$ & $-5.1e2$ & $-6.9e2$ & $0.$ & 
$5.7e2$ & $6.7e2$ & $-3.7e3$ & $-4.4e3$ & 
$2.2e3$ & $4.1e3$ & $1.4e3$ &   & 
  &   \\[1pt] 
$a_{hd}$ & $1.2e3$ & $4.3e2$ & $0.$ & 
$-8.1e2$ & $-1.4e3$ & $-1.3e2$ & $0.$ & 
$1.4e3$ & $-6.9e2$ & $-7.2e2$ & $0.$ & 
$6.7e2$ & $7.3e2$ & $3.3e3$ & $3.6e3$ & 
$4.2e2$ & $-2.9e3$ & $1.6e2$ & $1.1e3$ & 
  &   \\[1pt] 
$a_{he}$ & $-2.8e4$ & $-4.6e3$ & $-1.1e2$ & 
$9.0e3$ & $4.6e2$ & $-1.6e2$ & $23.$ & 
$-4.5e2$ & $2.5e2$ & $2.4e2$ & $-96.$ & 
$-1.7e2$ & $-3.0e2$ & $-2.5e4$ & $-3.2e4$ & 
$4.5e3$ & $1.7e4$ & $2.3e3$ & $-2.1e3$ & 
$3.2e4$ &   \\[1pt] 
$a_{W}$ & $7.7$ & $4.5$ & $0.$ & 
$-4.2$ & $0.$ & $0.$ & $0.$ & 
$0.$ & $0.$ & $0.$ & $0.$ & 
$0.$ & $0.$ & $6.3$ & $-1.7$ & 
$0.$ & $0.8$ & $0.$ & $0.$ & 
$1.4$ & $2.6$ \\[1pt] 
  & $a_{W\!B}$ & $a_{h}$ & $a_{ll}^s$ & 
$a_{ll}^t$ & $a_{lq}^s$ & $a_{lq}^t$ & $a_{le}$ & 
$a_{qe}$ & $a_{lu}$ & $a_{ld}$ & $a_{ee}$ & 
$a_{eu}$ & $a_{ed}$ & $a_{hl}^s$ & $a_{hl}^t$ & 
$a_{hq}^s$ & $a_{hq}^t$ & $a_{hu}$ & $a_{hd}$ & 
$a_{he}$ & $a_{W}$ \\[1pt] 
\end{tabular} }
\caption{\label{table:M}
The elements of the matrix ${\mathcal M}$. Since it is a symmetric matrix we do not list the 
redundant elements. The matrix is equal to the numbers listed above times $10^{12} ({\rm GeV})^4$.
We abbreviate the powers $10^n$ as $en$ to save space.}
\end{sidewaystable} 
\addtolength{\tabcolsep}{0.5pt}

The numerical values of the coefficients $\hat{a}_i$ and ${\mathcal M}_{ij}$
depend on both the experimental values and the SM predictions.
Should any of the SM input parameters change in the future, this affects the 
best fit values $\hat{a}_i$ in Eq.~(\ref{eq:main2}), but not the matrix ${\mathcal M}$.
The matrix ${\mathcal M}_{ij}$ only depends on the sizes of errors for different
measurements. Thus, ${\mathcal M}$ would change if experimental precision
improves in the future. The central, or best fit, values  $\hat{a}_i$ depend on all quantities:
central values of experiments, the errors, and the SM predictions.

At tree-level, SM predictions depend on well-measured quantities. However
the strong coupling constant, the top mass and the Higgs mass all enter 
at one-loop order. The least known of the three is the Higgs mass and it
is interesting to know how the predictions change as the Higgs mass is varied.
It is very easy to incorporate changes of the Higgs mass with respect to the
assumed reference value of $113~GeV$ in Ref~\cite{erler+langacker}, which we use
for the SM predictions. The dominant contribution from the Higgs are corrections to
gauge boson propagators and can be incorporated by
shifting the S and T parameters~\cite{STU}.  Higgs mass different than  
its reference value shifts the best fit value $\hat{a}_i$ as follows
\begin{equation}
  \delta \hat{a}_{WB}\approx \frac{\alpha}{48 \pi s c v^2} \log\left(\frac{m_h^2}{m_{h,ref}^2}\right), 
  \quad \delta \hat{a}_{h}\approx \frac{3 \alpha}{8 \pi c^2 v^2} \log\left(\frac{m_h^2}{m_{h,ref}^2}\right),
\end{equation}
where only the leading logarithm in the Higgs mass is kept~\cite{STU}. This is a good
approximation as can be verified by comparing with the exact one-loop result given in
Ref.~\cite{Marciano:1980pb}.

 \section{$S$ and $T$ parameters}
 \label{app:ST}
In our procedure, the $S$ and $T$ parameters correspond to 
$a_{W\!B}$ and $a_h$ as
$$S=\frac{4scv^2a_{W\!B}}{\alpha},\quad T=-\frac{v^2}{2\alpha} a_h.$$
Setting all $a_i$, but $a_{W\!B}$ and $a_h$, to zero in Eq.~(\ref{eq:main2}), we get
\begin{eqnarray}
\chi^2&=&\chi^2_0+
(a_{W\!B}\quad a_h)
\left(\begin{array}{cc}9.1\, 10^{16}&2.4\, 10^{16}\\2.4\, 10^{16}&7.9\, 10^{15}\end{array}\right)
\left(\begin{array}{c}a_{W\!B}\\a_h\end{array}\right)+1.5\, 10^8
a_{W\!B}-2.3 \,10^7 a_h\nonumber\\
&=&\chi^2_0+(S\quad T)\left(\begin{array}{cc}5.4\, 10^2 &-4.8\,  10^2\\ -4.8 \, 10^2&5.3\, 10^2\end{array}\right)
\left(\begin{array}{c}S\\ T \end{array}\right)+12. \, S + 5.9 \, T.
\label{eq:chiST}
\end{eqnarray}
A simultaneous fit to $S$ and $T$ using the above equation gives
$$S=-0.08\pm0.10, \quad T=-0.08\pm0.10.$$
The 1$\sigma$ error quoted above is obtained by projecting
the $\Delta \chi^2=1$ ellipse onto the corresponding axis. The 
plot of the 90\% confidence level contour is presented in Fig.~1. Comparing our results with
the result in Ref.~\cite{erler+langacker}, Figure~10.3, shows good agreement.
The results for $S$ and $T$ in Ref.~\cite{erler+langacker} do not include LEP2 measurements,
which indicates that LEP2 results do not significantly affect the bounds on the $S$ and
$T$ parameters.
\begin{figure}
\begin{center}
 \includegraphics[width=3.5in]{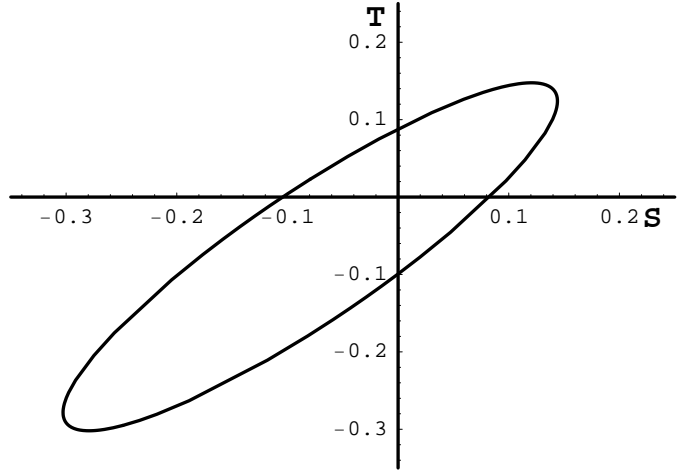}
\end{center}
\caption{Allowed region for $S$ and $T$ at 90\% confidence level
obtained using Eq.~(\ref{eq:chiST}).}
\end{figure}

 \section{Bounds on $Z'$ bosons}
 \label{app:Zprime}
Theoretical and experimental constraints on a color singlet neutral gauge boson
are discussed in Ref.~\cite{Zprime}. The SM gauge group is extended to include
a $U(1)_Z$ factor with $Z'$ its corresponding gauge boson. If the $Z'$
is heavier than the electroweak breaking scale, we can integrate it
out and obtain the following effective Lagrangian:
\begin{eqnarray}
 \Delta \mathcal{L}&=&-\frac{1}{2M_{Z'}^2}g_Z^2z_H^2|\phi^\dag
 D_\mu\phi|^2-\sum_{ff'}\frac{1}{4M_{Z'}^2}\frac{1}{1+\delta_{ff'}}g_Z^2
 z_f z_{f'} (\bar f\gamma^\mu f)(\bar f'\gamma_\mu f')
 \nonumber\\
 &&-\sum_f\left[\frac{i}{4M_{Z'}^2}g_Z^2 z_f z_H (\bar f\gamma^\mu
 \bar f )(\phi^\dag D_\mu\phi)+{\rm h.c.}\right],
 \label{zplagrangian}
\end{eqnarray}
where $M_{Z'}$ is the mass of $Z'$, $g_Z$ is the coupling constant for
$U(1)_Z$, and $z_H$ and $z_f$ are the $U(1)_Z$ charges for the Higgs
doublet and fermions.  These charges satsify \cite{Zprime}
 \begin{equation}
 z_l=-3 z_q,\quad z_e=-3 z_q-z_H,\quad z_u=z_H+z_q,\quad
 z_d=z_q-z_H.
 \label{charges}
 \end{equation}
Assuming electromagnetic strength for the Higgs-$Z'$ and fermions-$Z'$ couplings one obtains
 \begin{equation} z_H g_Z=g s,\quad z_q g_Z=\pm g
 s/3. \label{ewstrength}
 \end{equation}

The authors of Ref.~\cite{Zprime} considered three experimental
constraints on $M_{Z'}$: the bound on the $T$ parameter
implies a bound $M_{Z'}>0.9~\mbox{TeV}$; the total decay width of
the Z boson, $\Gamma_Z$, gives a bound $M_{Z'}>1.2~\mbox{TeV}$ for the
 $z_q g_Z=-g s/3$ case; and the left-right asymmetry of the
electron, $A_e$, gives a bound $M_{Z'}>1.0~\mbox{TeV}$ for the 
$z_q g_Z=+g s/3$ case. All the above bounds are given at 95\%
confidence level.

The T parameter corresponds to the coefficient of
the first term in Eq.~(\ref{zplagrangian}):
 \begin{equation}
 T=\frac{v^2}{4\alpha M_{Z'}^2}z_H^2 g_Z^2.
 \end{equation}
If we use the same value as in Ref.~\cite{Zprime}, $T=-0.02\pm0.13$,
we get the same bound for $M_{Z'}$. If we consider individual
measurements $\Gamma_Z$ and $A_e$, and use the results described in
Section~\ref{sec:Calculations}, we reproduce the other two bounds.

It is interesting to compare these bounds with a global fit to all data.
Using our formula Eq.~(\ref{eq:main2}), it takes little
effort to obtain a constraint that incorporates simultaneously all operators
in the effective Lagrangian (\ref{zplagrangian}). Using
Eqs.~(\ref{zplagrangian}) and (\ref{charges}) 
we obtain all the non-zero coefficients $a_i$:
\begin{equation}
a_h=-2 z_H^2 \beta,\quad a_{hf}=-z_H z_f\beta,\quad
a_{ff'}=-z_fz_{f'}\beta,
\end{equation}
where $\beta=g_Z^2/(4M_{Z'}^2)$.
(Cases such as $hf=hq$ are understood to be $a_{hq}^s$ and so on.)
 Substituting these coefficients in Eq.~(\ref{eq:main2})
and imposing Eq.~(\ref{ewstrength}), we obtain
$\chi^2$ as a function of $M_{Z'}$. We then find the bounds for $M_{Z'}$
at 95\% confidence level:
\begin{eqnarray}
M_{Z'}&>&2.2 \mbox{TeV }\qquad(z_q g_Z=\frac{g s}{3}),\\
M_{Z'}&>&2.4 \mbox{TeV }\qquad(z_q g_Z=- \frac{g s}{3}),
\end{eqnarray}
which are about twice as large as the bounds given in Ref.~\cite{Zprime}.

\section{A brief introduction to Mathematica packages.}
\label{app:program}
Numerical calculation of  $\chi^2$ was done using
Mathematica~\cite{Wolfram}. Our code can be obtained at
\textit{http://pantheon.yale.edu/$\sim$zh22/ew.html} or from the
authors. We provide two Mathematica no tebooks:
\textit{ew\_chi2\_calculations.nb} and
\textit{ew\_chi2\_results.nb}. The second notebook spares readers from retyping our 
results by giving the $\chi^2$ distribution in the form in Eq.~(\ref{eq:main2}).
It also gives the corresponding errors and the correlation matrix for the
coefficients of operators. For those who want to customize our calculation to better suit 
their purposes, the first notebook contains all the inputs, formulas,
and calculations. We briefly describe the structure of the program
below. More details are supplied in the comments inside the program.

The notebook
\textit{ew\_chi2\_calculations.nb} is coded in the following order: options,
input parameters, measurements, theoretical predictions, and the
calculation of $\chi^2$.

Three options have been implemented. First  option turns on or off
the initial state radiative corrections for fermion pair production at LEP2. (The
radiative corrections for $e^+e^-\rightarrow e^+e^-$ channel have not
been incorporated so far.)  Second option toggles if  the NuTeV result
is included in the calculation of $\chi^2$. The last one controls the presence
of second order terms in the coefficients of four-fermion operators at LEP2,
see Sec.~\ref{sec:chi2}.

The input parameters, experimental values, and SM predictions are given next.
Should any of these numbers change in the future, one needs to modify the program
accordingly.

Next, the deviations from the SM are calculated. All formulas
discussed in Sec.~\ref{sec:Calculations} can be found there. The
predictions are presented as the SM values plus corrections proportional to
 $a_i$. These predictions are then used to calculate the $\chi^2$ distribution. We
have split the total $\chi^2$ to track the contributions from
different measurements. 

\end{document}